\documentclass[11pt]{article}

\usepackage{a4wide}

\title{Instability criterion for oblique modes in stratified circular 
Couette flow}
\author{ C. Normand\\
Institut de Physique Th\'eorique, CNRS, URA 2306\\
CEA, IPhT, 91191 Gif-sur-Yvette Cedex, France.\\
contact: christiane.normand@cea.fr}

\begin{document}

\maketitle

\section*{Abstract}
An analytical approach is carried out that provides an inviscid stability criterion 
for the strato-rotational instability (in short SRI) occurring in a Taylor-Couette system.
The control parameters of the problem are the rotation ratio $\mu$ and the radius 
ratio $\eta$. The study is motivated by recent experimental \cite{legal} 
and numerical \cite{rudi, rudi2} results reporting the   
existence of unstable modes beyond the Rayleigh line for centrifugal 
instability ($\mu =\eta^2$). The modified Rayleigh criterion for stably 
stratified flows provides the instability condition, $\mu<1$, while in experiments  
unstable modes were never found beyond the line $\mu=\eta$. Taking into account  
finite gap effects, we consider non axisymmetric perturbations with 
azimuthal wavenumber $l$ in the limit $l Fr<<1$, where $Fr$ is the Froude number. 
We derive a necessary  condition for instability : $\mu < \mu^*$ where $\mu^*$,
a function of $\eta$, takes the asymptotic values, $\mu^* \to 1$ in the narrow 
gap limit, and $\mu^* \to 2\eta^2(1+\eta)$ in the wide gap limit,
in agreement with recent numerical findings. 
A stronger condition, $\mu<\eta$, is found when $\eta>0.38$, in 
agreement with experimental results obtained for $\eta=0.8$.
Whatever the gap size, instability is predicted for values of $\mu$ larger than
the critical one, $\mu_c=\eta^2$, corresponding to centrifugal instability.

\section{Introduction} 
Despite its long-lasting study since the work of Taylor \cite{taylor}, 
the stability of cylindrical Couette flow remains a vivid research area 
attracting many investigators. A survey of the literature on the topic can be found
in \cite{tagg}. The flow occurs in the annular gap between two
concentric cylinders of radii $R_1<R_2$ rotating independently at the
angular velocities $\Omega_1$ and $\Omega_2$, respectively. The control 
parameters are the radius ratio $\eta=R_1/R_2$ and the rotation ratio 
$\mu=\Omega_2/\Omega_1$. Inside the gap, the angular velocity profile of the laminar  
steady flow is $\Omega(r)$. Its linear stability with respect to axisymmetric 
perturbations is governed, in the inviscid limit, by the Rayleigh criterion  
\cite{lord} : $d(r^2\Omega)^2/dr>0$.
A generalized Rayleigh criterion for non-axisymmetric centrifugal instabilities
has been derived for a free axisymmetric vortex using a large axial
wavenumber WKB approximation \cite{biga}. The instability takes the form of
a spatially oscillating mode localized between two turning points where it 
matches with the exponentially decaying 
solutions outside. The generalization to take into 
account a background rotation and a stable stratification did not reveal  
fundamental changes in the Rayleigh criterion. This contrasts with what was 
found for bounded flows \cite{mole, dubru, umu}. In these studies, it was shown 
that a stable stratification drastically changes the stability criterion,
showing a strong analogy with the modified Rayleigh criterion obtained
when a magnetic field parallel to the cylinders' axis is present.
In the hydromagnetic case \cite{chandra, veli}  
the condition for stability with respect to axisymmetric perturbations 
is : $d(\Omega^2)/dr>0$. The presence of a stable stratification has a
similar effect on the Rayleigh criterion though its validity is in that  
case restricted to non axisymmetric perturbations \cite{mole, dubru, umu}.

In the astrophysical context, the magneto-rotational instability (MRI) was  
recognized as a potential source of turbulence in accretion disks \cite{balbus},   
especially Keplerian disks, with angular velocity $\Omega(r) \sim r^{-3/2}$, 
unable to sustain centrifugal instability according to the standard
Rayleigh criterion. The MRI mechanism is robust, it occurs in bounded as well
as unbounded flows and for compressible or incompressible fluid.
The capacity of the strato-rotational instability (SRI) to trigger turbulence
in Keplerian disks raises the open question of what are the appropriate 
radial boundary conditions at the edges of the disk. The existence of SRI has been 
demonstrated in the inviscid limit for perturbations that satisfy no normal  
flow conditions on the channel walls. In the viscous case, both the no-slip 
\cite{rudi} and stress-free \cite{dubru} conditions give rise to instability. 
Mixed boundary conditions in which no normal flow is imposed on one side of the  
channel and zero pressure on the other side were not able to sustain growing 
modes \cite{umu}. The effect of a vertically varying stratification
was also examined \cite{umu} showing that SRI persists in that case.

Beside possible applications in geophysics and astrophysics, SRI was  
studied in the laboratory.
Until very recently the conclusion of experimental studies was that
a stable vertical stratification stabilizes the flow
\cite{chen, boub, caton}. The experimental 
evidence of the strato-rotational instability (SRI) in a Taylor-Couette system
was definitely assessed in \cite{legal}. Moreover, the values of the control  
parameters for which instability occurs were found in good agreement with  
numerical predictions \cite{rudi} that yields the condition : $\mu < \eta$, for instability.

The aim of the present contribution is to derive an instability criterion by an 
entirely analytical analysis that improves previous studies achieved in the small  
gap limit \cite{mole, dubru, umu}. When curvature effects are neglected the problem 
under consideration reduces to the stability of a stratified plane Couette
flow rotating at constant angular velocity $\Omega$. In cartesian coordinates $(x, y, z)$
the flow velocity is ${\bf V}=(0, Sx, 0)$ where $S$ is the constant shear. 
The condition for instability in the stratified case is $S/2 \Omega<0$, while the standard
Rayleigh-Pedley criterion gives : $S/2 \Omega<-1$, when there is no stratification. 
The substitution $S \to r \Omega^{'}$, is often used to deduce the modified Rayleigh 
criterion for stratified flows with curved streamlines : $d(\Omega^2)/dr<0$. 
When applied to the circular Couette flow, the instability criterion
simply yields : $\mu<1$. 
However, in experiments \cite{legal} achieved for a finite value of the gap,
unstable modes were never found beyond the line $\mu =\eta$, in agreement 
with previous numerical results \cite{rudi}. Quite recently, the situation 
has changed since numerical calculations \cite{rudi2} have shown that the 
stability line has a more complicated dependence on $\eta$. In the narrow gap 
limit, the stability limit was found beyond the line $\mu=\eta$, while in the 
wide gap limit it was found in between the lines $\mu=\eta^2$ and $\mu=\eta$.
The narrowing of the instability range, when the gap size increases,
 is possibly due to curvature effects that were not
taken into account with sufficient accuracy in the transposition
$S \to r \Omega^{'}$. In the present contribution a more appropriate treatment
of curvature effects is carried out leading to
an instability criterion that involves the two 
parameters $\mu$ and $\eta$. Our stability results will be  
compared to the experimental ones \cite{legal} obtained for a value of the   
radius ratio $\eta=0.8$, and to the numerical ones \cite{rudi, rudi2} for three
values of the gap in the range $0.3\le \eta \le 0.78$.

\section{Stratified circular Couette flow}
In cylindrical coordinates $(r, \varphi, z)$ the velocity field in the basic  
state is ${\bf V}=(0, r \Omega(r),0)$ with the angular velocity given by \cite{rudi}
\begin{equation}
\Omega(r)=\Omega_1\left(\frac{A}{r^2} +B\right)=\Omega_1 \widehat \Omega(r) 
\quad \mbox{where} \quad A= R_1^2 \frac{1-\mu}{1-\eta^2}, 
\quad \mbox{and} \quad B = \frac{\mu -\eta^2}{1-\eta^2} \label{CCF}
\end{equation}
The fluid is assumed incompressible with a stable density stratification
along the vertical cylinders' axis ($\partial \rho/\partial z<0$). 

\subsection{Linearized equations for perturbations}
The governing linearized equations for the perturbed velocity
 ${\bf u}=(\tilde u, \tilde v, \tilde w)$,
the pressure and entropy perturbations, respectively $\tilde p$ and $\tilde h$ are given 
in \cite{mole}. In the study of stratified plane Couette flow \cite{dubru, umu}
it was particularly convenient to reduce the full set of equations to a set of 
coupled equations for the radial velocity component $\tilde u$ and
the pressure $\tilde p$. Following the same procedure, the perturbed quantities are sought 
in the form $(\tilde u, \tilde p)=\exp i(mz +l \varphi +\omega t)[(u(r), p(r)]$ where $m$ 
and $l$ are respectively the axial and azimuthal wavenumbers, these notations are consistent  
with \cite{mole} but not with \cite{rudi} where $m$ has a different meaning. 
The radial dependencies $u(r)$ and $p(r)$ satisfy
\begin{eqnarray}
\sigma D_* u + \frac{l}{r} Z u &=&-i \frac{G}{K} p \label{Duc}\\
(\sigma^2 -2\Omega Z) u - 2 i \frac{l}{r} \Omega p &=& -i \sigma Dp, \label{Dpc}
\end{eqnarray}
where $D\equiv \partial/\partial r$ and $D_*=D+1/r$. Here, $\sigma=\omega-l \Omega$, 
and $Z=2\Omega + r \Omega^{'}$. 
The quantities $K$ and $G$ are given below
\begin{equation}
K=\sigma^2-N^2, \quad  \mbox{and} \quad G=\sigma^2 m^2 +\frac{l^2}{r^2} K
\end{equation}
where $N$ is the Brunt-V\"ais\"al\"a frequency. For circular Couette flow with 
angular velocity (\ref{CCF}), the value of $Z=2\Omega_1 B$,  
is a constant. Thus, the Rayleigh discriminant, $\Phi=2\Omega Z$, is proportional to 
$\Omega$ and it can be written :
\begin{equation}
\Phi=4\Omega_1^2 B {\widehat \Omega} \equiv \Omega_1^2 {\widehat \Phi} \label{ray}
\end{equation}
The system (\ref{Duc})-(\ref{Dpc}) can be reduced to a second order differential 
equation for the radial component $u$, as in \cite{biga, mole}. However, such
a formulation leads to tractable results only in a few cases, 
for instance in the large axial wavenumber limit ($m>>1$) 
or the small gap limit ($\eta \to 1$). The present stability analysis
will closely follow the procedure done in \cite{umu} for the plane Couette flow. 
The main difference is the consideration of curvature effects. However, the 
asymptotic analysis \cite{umu} based on the small azimuthal wavenumber assumption  
cannot be transposed directly in the circular case where $l$ takes integer 
values : $l=1,2,\cdots$. In the circular case, we shall assume the following scalings
\begin{equation}
\omega = l\Omega_0 \bar \omega, \quad \Omega= \Omega_0 \bar \Omega, \quad \Rightarrow \quad
\sigma = l\Omega_0 \bar \sigma, \quad K=-N^2 (1- l^2 Fr^2 \bar \sigma^2)
\end{equation}
where $\Omega_0=(\Omega_1+\Omega_2)/2$, is the mean angular velocity and $Fr=\Omega_0/N$,
the Froude number. The asymptotic value $K \to -N^2$ obtained in the limit
$l<<1$ \cite{umu} is recovered here in the limit $l Fr<<1$,  that can be reached
either when $l<<1$ and $Fr \approx {\cal O}(1)$
or $l\approx{\cal O}(1)$ and $Fr<<1$. Another important assumption concerns the term 
$H=\sigma^2 -2\Omega Z$ in Eq. (\ref{Dpc}) that was approximated in \cite{umu} by 
$H \to -\Phi$ where $\Phi$ the Rayleigh discriminant given in Eq. (\ref{ray}) is now
written $\Phi=\Omega_0^2 \bar \Phi$ leading to $H=\Omega_0^2(l^2 \bar \sigma^2 - \bar \Phi)$.
Noticing that $\bar \Phi$ can takes values as large as  
$4\Omega_1^2/\Omega_0^2$, the assumption $l<<1$ will be replaced by the weaker 
condition : $l<l_{max}$ with $l_{max}=2\Omega_1/\Omega_0$. To illustrate our purpose, 
this gives the following constraints : $l<2.2$ for $\mu=0.8$ and  $l<3$ for $\mu=0.3$.  
Thus, the lower is $\mu$, the higher is the allowed value of $l$. In the following,  
we shall consider the limit $l Fr<<1$ with $l\le l_{max}$. This leads to   
simplifications in Eqs. (\ref{Duc})-(\ref{Dpc}) that are listed below
\begin{equation}
K \to -N^2, \quad H \to -\Phi, \quad \frac{G}{K} =-l^2\left(\bar m^2 \bar \sigma^2 -
\frac{1}{r^2} \right) = -l^2g(r)
\end{equation} 
where $\bar m=m Fr$.

It must be stressed that the generalization of the Rayleigh criterion \cite{biga}
performed in the limit $m>>1$, is not equivalent to the approach described here. 
The scaling used in \cite{biga} for the frequency, suitable in describing
inertial-gravity waves, could explain why SRI was not found in this study. 

\subsection{Governing equations in the limit $l Fr<<1$ and $l<l_{max}$}
The perturbations $u$ and $p$ are scaled according to
\begin{equation}
 u = l \bar u, \quad \mbox{and} \quad  p=\Omega_0 \bar p
\end{equation}
and substituted in (\ref{Duc})-(\ref{Dpc}) leading to the following set of equations : 
\begin{eqnarray}
\bar \sigma \ D_* \bar u + \frac{\bar Z}{r} \bar u &=& i\ g\ \bar p \label{Du1c}\\
\bar \Phi \bar u + 2 i \frac{\bar \Omega}{r}  \ \bar p &=& i\ \bar \sigma \ D\bar p \label{Dp0c}
\end{eqnarray}
where $\bar \Omega= \Omega/\Omega_0$ and $\bar Z=Z/\Omega_0$. We have reported in 
Appendix A the equations that should be solved  for $l\ge l_{max}$ when the 
limit $H \to -\Phi$ is no longer valid. 

Eqs. (\ref{Du1c})-(\ref{Dp0c})
are the circular version of those obtained for a plane Couette flow \cite{umu}, they 
constitute a set of coupled equations provided $\Phi \ne 0$.  
It was shown in \cite{umu} that elimination of $\bar u$ leads to a 
simple equation for $\bar p$, its two independent solutions being exponential functions. 
Then substituting $\bar p$ in (\ref{Dp0c}) gives immediately
the expression for $\bar u$. Finally, satisfaction of the conditions $\bar u=0$, on the 
two boundaries confining the flow leads to the instability criterion. The same 
procedure is applied here to the cylindrical case. After some calculations, the first    
step gives the equation satisfied by $\bar p$ 
\begin{equation}
\bar \sigma \left[ D^2 \bar p + \left(\frac{1}{r}-\frac{\Phi^{'}}{\Phi}\right) D \bar p - 
\bar m^2 \bar \Phi \bar p \right] + 2 \frac{\bar \Omega}{r} \left[{\partial \over \partial r}
Log\left(\frac{\bar \Phi}{\bar \Omega}\right)\right] \bar p = 0 \label{D2p0c}
\end{equation}
In the general case, the solutions of (\ref{D2p0c}) cannot be expressed in terms of simple 
analytical functions. However, for circular Couette flow 
with angular velocity given by (\ref{CCF}) the Rayleigh discriminant is proportional 
to $\bar \Omega$ and the derivative of the logarithmic term in (\ref{D2p0c}) vanishes. 
This remarkable feature occurs for angular velocity profiles of the type $\Omega(r)=C r^x +D$,
where $C$ and $D$ are constant coefficients.
This allows considerable simplification of Eq. (\ref{D2p0c}) that,
provided $\bar \sigma \ne 0$, becomes 
\begin{equation}
D^2 \bar p + \left(\frac{1}{r}-\frac{\Omega^{'}}{\Omega}\right) D \bar p - 
\bar m^2 \bar \Phi \bar p =0. \label{pbar}
\end{equation}
It should  be noticed that the term in the left-hand-side of Eq. (\ref{pbar}) 
is reminiscent of the potential vorticity describing the content of a Rossby wave 
as mentioned in \cite{umu}.
In the small gap limit, $\eta \to 1$, with $r/R_1=1+(1-\eta)(x/\eta)$ and $\Omega$ constant, 
the term in front of $D\bar p$ disappears from (\ref{pbar}) thus recovering the solution
$\bar p = \exp(\pm \bar m {\sqrt {\bar \Phi}} x)$ found in \cite{umu} for the  stratified rotating 
plane Couette flow when the Rayleigh criterion for stability, $\Phi>0$, is satisfied. 
Then, introducing  $\hat p=\bar p/{\sqrt {\bar \Omega}}$, it satisfies
\begin{equation}
{\cal L}_0 \hat p + \left[\frac{A}{\widehat \Omega r^4}\left(2-\frac{3A}{\widehat \Omega r^2}\right) - 
4\widehat m^2 B \widehat \Omega \right] \hat p  = 0 \label{Lhatp}
\end{equation}
with the differential operator ${\cal L}_0 = D_*D$ accounting for curvature effects.
In Eq. (\ref{Lhatp}) use has been made of the identity $\bar m^2 \bar \Phi=\widehat m^2 \widehat \Phi$,
where $\widehat m=m\Omega_1/N$ and $\widehat \Phi=4B \widehat \Omega$.
We shall assume that $\widehat m$ is large enough so that the term in $\widehat m^2$ 
dominates in Eq. (\ref{Lhatp}). Our analysis differs from previous studies  
that linearized the angular velocity : $\Omega(r) = \Omega(r_0)+(r-r_0)\Omega^{'} +\cdots$
around a mean radius $r_0$ as in the thin gap limit. Here, the 
exact expression of $\widehat \Omega(r)$ given in (\ref{CCF}) is substituted in (\ref{Lhatp}) 
that becomes
\begin{equation}
{\cal L}_0 \hat p - 4 \widehat m^2 B \left(\frac{A}{r^2}+B\right) \hat p  = 0. \label{L0p}
\end{equation}
When $A$ and $B$ have the same sign, the solutions of (\ref{L0p}) are the modified Bessel  
functions $I_n(\alpha r)$ and $K_n(\alpha r)$ of general order $n$ real positive with
$n= 2 \widehat m \sqrt{A B}$ and $\alpha=2 \widehat m B$. The expression for $\hat p$ is
\begin{equation}
 \hat p = A_0 I_n(\alpha r) + B_0 K_n(\alpha r) \label{hatp}
\end{equation}
where the unknown coefficients $A_0$ and $B_0$ will be determined by the satisfaction 
of the boundary conditions on $\bar u$. Substituting (\ref{hatp}) in (\ref{Dp0c}) and  
using the relation between the Bessel functions and their derivatives, one gets
$\bar u= {\hat u}/\sqrt {\bar \Omega}$ with
\begin{equation}
\hat u =A_0 \ U_1(r) +B_0 \ U_2(r)
\end{equation}
Using vector notations, the two functions $U_1(r)$ and $U_2(r)$ are considered as
components of the bidimensional vector ${\bf U}=(U_1, U_2)$. Similarly, introducing
the vector ${\bf \Psi}_{\nu}=(I_{\nu}(\alpha r), K_{\nu}(\alpha r))$ for $\nu \in (n, n-1)$
gives the expression for $\bf U$
\begin{equation}
{\bf U}=2 \frac{{\bar \Omega}}{r} {\bf \Psi}_n + \bar \sigma \left( \left(\frac{n}{r}-
\frac{\Omega^{'}}{2\Omega}\right) {\bf \Psi}_n \mp \alpha  {\bf \Psi}_{n-1}\right)
\end{equation}
In the following, we shall neglect the terms in $\Omega^{'}/\Omega$ that appear in factor
of $\bar \sigma$, which are small compared to the terms proportional to $n$ or $\alpha$ 
which behave like $\widehat m$.

Satisfaction of the boundary conditions $\bar u(R_1)=\bar u(R_2)=0$ provides the algebraic 
system 
\begin{equation}
{\bf M}{\bf A}=0 \label{AS1}
\end{equation} 
with the vector ${\bf A}=(A_0,B_0)$ and the elements of
the matrix ${\bf M}$ given by $M_{ij}=U_j(R_i)$. Their expressions are easily deduced from
\begin{equation}
{\bf U}(R_i)=\frac{\lambda_i}{R_i} {\bf \Psi}_n^i \mp \alpha {\sigma}_i {\bf \Psi}_{n-1}^i
\end{equation}
where ${\sigma}_{i}=\bar \omega - \bar \Omega_i$ and 
$\lambda_i=2 \bar \Omega_i + n {\sigma}_{i}$ with $\bar \Omega_i =\bar \Omega(R_i)$
for $i=1,2$. We have also
introduced ${\bf \Psi}_n^i = (I_n^{(i)}, K_n^{(i)})$ where the abbreviations
$I_n^{(i)}=I_n(\alpha R_i)$ and $K_n^{(i)}=K_n(\alpha R_i)$ have been used. 

\subsection{Dispersion relation}
The vanishing of the determinant associated to the algebraic system in Eq. (\ref{AS1}) 
leads to the dispersion relation
\begin{equation}
\frac{\lambda_1}{R_1}\frac{\lambda_2}{R_2} S_n - \alpha^2 {\sigma}_{1}{\sigma}_{2} S_{n-1}
+\alpha {\sigma}_{2}\frac{\lambda_1}{R_1}C_{12} - \alpha { \sigma}_{1}\frac{\lambda_2}{R_2} C_{21}=0
\label{dispCC}
\end{equation}
where the following quantities have been introduced :
\begin{equation}
S_n  =  I_n^{(1)}K_n^{(2)}-I_n^{(2)}K_n^{(1)} \quad \mbox{and} \quad
C_{ij}  =  I_n^{(i)}K_{n-1}^{(j)}+I_{n-1}^{(j)}K_n^{(i)}\quad \mbox{for}\quad (i,j)\in (1,2).
\label{SC}
\end{equation}
To simplify the calculations we shall use the asymptotic expansions at large  
arguments of the modified Bessel functions $I$ and $K$. At the lowest order, 
the asymptotic behaviors of expressions (\ref{SC}) are :
\begin{equation}
S_n =S_{n-1} \approx \frac{\sinh(\alpha_1 -\alpha_2)}{\sqrt{\alpha_1\alpha_2}}  \quad \mbox{and} \quad C_{12}=C_{21} 
\approx \frac{\cosh(\alpha_1 -\alpha_2)}{\sqrt{\alpha_1\alpha_2}} 
\end{equation}
Introducing $q=\alpha(R_2-R_1)$, the dispersion relation (\ref{dispCC}) reads
\begin{eqnarray}
T_1 - T_2 \coth q = 0&  & \label{disp2}\\
\mbox{where} \quad T_1=\left(\frac{\lambda_1 \lambda_2}{R_1 R_2} - \alpha^2 {\sigma}_{1}{\sigma}_{2}\right) 
\quad \mbox{and}& & \quad T_2 =  \alpha \left({\sigma}_{2}\frac{\lambda_1}{R_1}
 - {\sigma}_{1}\frac{\lambda_2}{R_2}\right)
\label{T12}
\end{eqnarray}
are second order polynomials in $\bar \omega$ that will be expressed
in terms of the rate of shear ${\cal S}$ given below:
\begin{equation}
 {\cal S}=\frac{\mu-1}{\mu+1}<0, \quad \Longrightarrow \quad \left\{\begin{array}{c}
 \sigma_1=\bar \omega -1 + {\cal S}\\
\sigma_2=\bar \omega -1 - {\cal S}  \end{array} \right.
\end{equation}
Doing the change of variable $\hat \omega = \bar \omega-1$,
expressions (\ref{T12}) becomes
\begin{eqnarray}
T_1 & = &  X ({\hat \omega}^2 - {\cal S}^2)  + 
\frac{4n}{R_1 R_2}(\hat \omega +{\cal S}^2) +  \frac{4(1-{\cal S}^2)}{R_1 R_2}\label{t1}\\
T_2 & = &  {\alpha \over R_1}\left\{2\left[(1-\eta) (\hat \omega +{\cal S}^2)
-(1+\eta){\cal S}(1+\hat \omega)\right] + n (1-\eta)({\hat \omega}^2 - {\cal S}^2)\right\}
 \label{t2}
\end{eqnarray}
The quantity $X$ involved in expression (\ref{t1}) for $T_1$   
can be expressed in terms of the control parameters, $\mu$ and $\eta$, as follows 
\begin{equation}
X =\frac{n^2}{R_1 R_2}-\alpha^2\equiv 4 {\widehat m}^2 B \frac{\eta-\mu}{1-\eta}
\label{X}
\end{equation}

\subsection{Instability criterion}
It is worth while noticing that the sign of the quantity $\eta-\mu$ which appears in  
expression (\ref{X}) for $X$ seems determinant for the stability of the system   
according to what has been observed both in experiments \cite{legal} and in  
 numerical computations \cite{rudi}. To check whether our calculations 
support this finding we shall determine the roots of the dispersion relation 
(\ref{disp2}). Instability could occur if there is a pair of complex conjugate roots,
one with negative imaginary part corresponding to instability growth, the other to
decay. The calculation of the discriminant and the determination of its sign is 
facilitated by introducing the following quantities 
\begin{equation}
t_0=X - n (1-\eta) Q, \quad \mbox{with} \quad Q=\frac{\alpha}{R_1} \coth q, \label{Q}
\end{equation}
and 
\begin{equation}
 t_1= \frac{4n}{R_1 R_2} - 2(1-\eta) Q, \quad \quad \quad
 t_2=2(1+\eta){\cal S}Q. \label{t12}
\end{equation}
The dispersion relation (\ref{disp2}) now reads 
\begin{equation}
{\hat \omega}^2 t_0 + {\hat \omega}(t_1+t_2)-{\cal S}^2(t_0 - t_1) +t_2 +\frac{4}{R_1 R_2}(1 -{\cal S}^2)=0
\end{equation}
and the associated discriminant is
\begin{equation}
\Delta = (t_1 + t_2)^2-4 t_0\left[-{\cal S}^2(t_0 - t_1) +t_2+\frac{4}{R_1 R_2}(1 -{\cal S}^2)\right]
\label{Del}
\end{equation}
Instability could occur if the discriminant is negative. To determine the 
sign of $\Delta$ two cases will be considered according to the sign of $t_0$.  
Expression (\ref{Q}) for $t_0$ shows that it is necessarily negative when $X<0$,
which occurs for $\mu>\eta$. On the opposite side, when $X>0$ for $\mu<\eta$, $t_0$  
can takes positive as well as negative values. The change of sign of $t_0$
occurs when
\begin{equation}
 X = n (1-\eta) Q \label{t0p}
\end{equation}
Replacing $X$, $n$ and $Q$ by their expressions in terms of $\mu$ and $\eta$, the 
above equation becomes 
\begin{equation}
  \coth q = b_0, \quad \mbox{with} \quad b_0 = \frac{(1+\eta)(\eta-\mu)}
 {(1-\eta)\sqrt{(1-\mu)(\mu-\eta^2)}} \label{up2}
\end{equation}
Eq. (\ref{up2}) can only be satisfied if $b_0>1$.
Writting that the square of $b_0$ is larger than unity, gives the 
following relation between $\mu$ and $\eta$
\begin{equation}
 2 (1+\eta^2)\mu^2 -\mu [(1+\eta^2)^2 + 4 \eta^2] + 2 (1+\eta^2)\eta^2>0 \label{poly}
\end{equation}
The left-hand-side of (\ref{poly}) is a second order polynomial in $\mu$ with coefficients 
depending on $\eta$. It takes a positive value for values of $\mu$ ranging outside the 
interval between the two roots $\mu_-=2\eta^2/(1+\eta^2)$ and $\mu_+=(1+\eta^2)/2$.
The root $\mu_+>\eta$, introduced when taking the square of $b_0$ in (\ref{up2})
is spurious and will not be considered. Thus, the meaningful condition for the satisfaction 
of $b_0>1$ is :  
\begin{equation}
\mu<\mu_- \quad \mbox{with} \quad \mu_-=\frac{2\eta^2}{(1+\eta^2)}
\end{equation}
Numerical values of $\mu_-$ are reported in Table 1 for different values of $\eta$.
The above results are summarized as follows. For $\mu>\mu_-$ and $b_0<1$, we have 
$t_0<0$, whatever the value of $q$.
For $\mu<\mu_-$, and $b_0>1$, one can write $b_0=\coth q_0$ and the sign of $t_0$ 
depends on the value of $q$. One gets 
\begin{equation}
t_0>0 \quad \mbox{when} \quad q>q_0 \quad \mbox{or}
 \quad t_0<0 \quad \mbox{when} \quad q<q_0. \label{t0pm}
\end{equation}
Considering separately the two cases, $t_0<0$ and $t_0>0$, we shall derive the conditions
for instability ($\Delta<0$) in the next sections.
 
\subsubsection{$t_0<0$}
It was shown in the previous section that $t_0$ can be negative for any value  
of $\mu$ but with the additional constraint $\coth q>b_0$ when $\mu<\mu_-$.
When $t_0<0$, a necessary condition for instability could be derived from the 
requirement that the term in factor of $t_0$ in Eq. (\ref{Del}) is negative
\begin{equation}
-{\cal S}^2(t_0 - t_1) +t_2+\frac{4}{R_1 R_2}(1 -{\cal S}^2)<0
\end{equation}
Replacing $t_0$,  $t_1$ and $t_2$ by their expressions (\ref{Q})-(\ref{t12}) 
and proceeding to some manipulations, one gets
\begin{equation}
-{\cal S}^2 X +{\cal S} Q\left[2 a+n{\cal S}(1-\eta)\right]+
\frac{4}{R_1 R_2}\left[(1 -{\cal S}^2) + n {\cal S}^2\right]<0 \label{t0n}
\end{equation}
where the positive quantity $a$ is expressed in terms of $\mu$ and $\eta$
\begin{equation}
a=1-{\cal S}+\eta (1+{\cal S})=\frac{2(1+\mu \eta)}{1+\mu} \le 2
\end{equation}
Then, expression (\ref{X}) for $X$ is substituted in the left-hand-side of Eq. (\ref{t0n})
that is written as the sum of two terms, leading to 
\begin{equation}
P_1 + {\cal S}^2 P_2 <0,  \label{p12}
\end{equation}
 with
\begin{eqnarray}
P_1& = &\alpha^2 {\cal S}^2 +2 a {\cal S}\frac{\alpha}{R_1} \coth q + \frac{4}{R_1 R_2} \label{p1}\\
P_2 & = & -\frac{(n-2)^2}{R_1 R_2} + (1-\eta)n\frac{\alpha}{R_1} \coth q \label{p2}
\end{eqnarray}
It is worth noticing that in the small gap limit ($\eta \to 1$) the term $P_2$ is  
irrelevant since the terms in $n$ do not exist and ${\cal S}^2<<1$. 
In this limit, when $R_1 \sim R_2$ and $a \sim 2$,  Eq. (\ref{t0n}) reduces to : $P_1<0$, 
where $P_1$ can be factorized, to get
\begin{equation}
\left(\alpha {\cal S} + \frac{2}{R_1} \coth \frac{q}{2}\right)
 \left(\alpha {\cal S} + \frac{2}{R_1} \tanh \frac{q}{2}\right)<0 \label{p1n}
\end{equation} 
thus recovering the condition for instability, ${\cal S}<0$, found for stratified rotating plane 
Couette flows. The instability occurs for values of $q$ that belong to an interval 
bounded by the two values, $q_-$ and $q_+$, deduced from
\begin{equation}
\tanh \frac{q_-}{2} =\gamma \frac{q_-}{2}, \quad \mbox{and} \quad \coth \frac{q_+}{2} =\gamma \frac{q_+}{2}
\quad \mbox{with} \quad \gamma=\frac{\eta \vert{\cal S}\vert}{(1 - \eta)}.
\end{equation}
For $\gamma>1$,  
the instability region is $q<q_+$ while for $\gamma<1$ the values of $q$ are in the 
interval $q_-<q<q_+$. It is found that $\gamma=1$ when $\mu=2 \eta-1$. 
For $\eta<0.5$, the value of $\gamma$ is lower than unity, whatever the value of $\mu$.

When curvature effects are present, the occurrence of instability could be
suppressed when the term $P_2$ is positive and ${\cal S}^2 P_2> \vert P_1\vert$.
As it will be intricate to derive a global condition for satisfaction of Eq. (\ref{p12})  
it will be replaced by a stronger constraint that consists to impose separately 
$P_1<0$ and $P_2 \le 0$.
Assuming that $(n-2)^2\approx n^2$ in Eq. (\ref{p2}), then $P_2$ will be negative
for values of $q$ satisfying the condition given below
\begin{equation}
 \coth q \le  \frac{b_1}{(1-\eta)} \quad \mbox{where} \quad
  b_1= \frac{n}{\alpha R_2}=
 \eta \left(\frac{1-\mu}{\mu -\eta^2}\right)^{1/2} \label{p2n}
\end{equation} 
The upper bound for $\coth q$ in (\ref{p2n}) has to be larger than unity, this occurs for
\begin{equation}
\mu \le \mu^* \quad \mbox{with} \quad \mu^* = \frac{\eta^2[1+(1-\eta)^2]}{\eta^2 +(1-\eta)^2}
 \label{ci1}
\end{equation}
For values of $\mu$ lower than $\mu^*$, $P_2$ will be negative for values of $q$ such 
that $q>q^*$ with $\coth q^*=b_1/(1-\eta)$. We have checked for some representative 
values of $\mu$ and $\eta$
that values of $q$ inside the interval $[q_-, q_+]$ are larger than $q^*$. Therefore, 
the two conditions $P_1<0$ and $P_2<0$ can be satisfied simultaneously.
The stability limit $\mu^*$ admits asymptotic values. In the narrow gap limit ($\eta \to 1$)
it is found that $\mu^* \to 1$, which is the stability line for plane Couette flows, while 
in the wide gap limit $\mu^* \to 2 \eta^2 (1+\eta)$. Moreover, the value of the gap size for 
which $\mu^*=\eta$ corresponds to $\eta=0.38$. Numerical values of $\mu^*$ are
reported in Table 1 for different values of $\eta$.
 
Eq. (\ref{ci1}) provides a necessary condition for instability. To be a sufficient  
one the positive term $(t_1+t_2)^2$ in expression (\ref{Del}) for the discriminant  
should not exceed the negative term. The term $(t_1+t_2)$ could  
even vanish if
\begin{equation}
\coth q =b_1 \frac{(1+\mu)}{(1-\mu \eta)} \equiv \coth q_{12} \label{csuf}
\end{equation}
The value $\coth q_{12}$ is consistent with the values of $\coth q$ allowed 
by Eq. (\ref{p2n}) if 
\begin{equation}
 \frac{(1+\mu)}{(1-\mu \eta)} \le \frac{1}{(1-\eta)} \quad  \Longrightarrow  \quad \mu<\eta \label{bsuf}
\end{equation}
When $\eta>0.38$, the condition, $\mu<\eta$, provides a stronger 
constraint than the condition $\mu<\mu^*$ coming from Eq. (\ref{ci1}). 

We have not found the specific value, $\mu=\mu_s$, for which the positive 
and negative terms exactly cancel in the expression for the discriminant
written in Eq. (\ref{Del}). Thus, the upper bound $\eta$ found in 
Eq. (\ref{bsuf}) is a low estimate of the exact value
$\mu_s$ which is more likely expected in the range $\eta \le \mu_s \le \mu^*$.

The above results are stricly valid for values of the rotation rate such that 
$\mu_- <\mu<\mu^*$. In that case, the constraints on the value of $q$, which are 
respectively $q \in [q_-, q_+]$ and $q>q^*$, can be satisfied simultaneously.  
When $\mu<\mu_-$, an additional constraint has to be satisfied : $q<q_0$,
as shown in Eq. (\ref{t0pm}). Instability could occur if there is an overlap 
of the two intervals $[q_-, q_+]$ and $[q^*, q_0]$ which requires 
that $q_0>q_-$. In that case instability occurs for $q \in [q_-, q_0]$.

\subsubsection{$t_0>0$}
In the previous section, when investigating the case $t_0<0$, 
the existence of instability was assessed for $\mu_- <\mu<\eta$ and 
$q \in [q_-, q_+]$. For $\eta^2 <\mu <\mu_-$, the instability 
is restricted to a narrower range of values of $q$  and its existence was 
not demonstrated in a systematic way. For $\mu <\mu_-$, we shall now 
investigate the case $t_0>0$ that could be more favorable to SRI.
 
When $t_0>0$, the terms in the left-hand-side of Eq. (\ref{Del}) are rearranged to read
\begin{equation}
\Delta = \left(t_1 + t_2 -2 {\cal S} t_0\right)^2 +4^2 (1-{\cal S}) t_0 t_3 \quad \mbox{with}
\quad t_3 =
{\cal S} \left(\frac{n}{R_1 R_2}-Q\right)-\frac{(1+{\cal S})}{R_1 R_2} \label{Dnew}
\end{equation}
In that case, the discriminant could be negative if $t_3$ is negative. 
The change of sign of $t_3$  occurs when
\begin{equation}
\coth q = \frac{1}{R_2}\left(\frac{n}{\alpha} + \frac{(1+{\cal S})}
{\alpha \vert {\cal S} \vert}\right)=b_1 +b_2  \label{t3n}
\end{equation}
The value of $\coth q$ in Eq. (\ref{t3n}) is the sum of two positive contributions.
The first contribution coincides with $b_1$ given in Eq. (\ref{p2n}), it
is independent of $\widehat m$ and for $\mu \le \mu_-$ it stands in the range 
: $1 \le b_1\le b_0$, 
 so that we can write $b_1=\coth q_1$.
The second contribution $b_2 \sim \alpha^{-1}$, that behaves like $\widehat m^{-1}$ 
 can be neglected for large values of $\widehat m$. In that case, $t_3$ will be 
negative for values of $q$ such that
\begin{equation}
\coth q \le b_1 \quad \mbox{or} \quad  q>q_1, \label{t3q}
\end{equation}
which is consistent with the condition : $\coth q \le b_0$, allowing for $t_0>0$.  
As soon as the value of $\widehat m$ decreases, the contribution $b_2$ 
increases until it reaches a value such that $b_1+b_2 = b_0$.
In that case, $t_3$ will be negative for values of $q$ corresponding to 
$\coth q \le b_0$ which exactly coincides with the condition ensuring 
$t_0$ is positive. Therefore, considering smaller values of $\widehat m$ 
will never give a stronger condition and moreover it will be contradictory
with the assumption leading to Eq. (\ref{Lhatp}). 

When $t_0>0$, the necessary condition for SRI is of the type
\begin{equation}
\coth q<\coth q_{max} \quad \mbox{with} \quad q_0 \le q_{max} \le q_1,
\end{equation}
meaning that instability might occur for $q>q_{max}$, the value of $q_{max}$
depending on the value of $\widehat m$. To find a sufficient
condition for SRI it will be argued as in the previous section where we have
looked for the value of $\coth q$ corresponding to the vanishing of 
the positive terms in the discriminant. In Eq. (\ref{Dnew}) it occurs when 
$t_1 +t_2=2 {\cal S} t_0$, 
that could be solved only if $t_1 +t_2<0$. Having determined in Eq. (\ref{csuf})  
the value of $\coth q=\coth q_{12}$ for which the sum $t_1 +t_2$ vanishes,
the sum will be negative for $\coth q>\coth q_{12}$
 or correspondingly $q<q_{12}$. As $\coth q_{12}>\coth q_{1}$, 
and consequently $q_{12}< q_{1}$, the simultaneous satisfaction of $q>q_{max}$ and 
$q<q_{12}$ is only possible when $q_{max}$ belongs to the interval 
$q_0<q_{max}<q_{12}$ which implies to consider values of $\widehat m$ not too 
large. In that case, instability will occurs for $q_{max}<q<q_{12}$. Hence,  
whatever the sign of $t_0$, the values of $q$ leading to instability are 
restricted to a limited band.

\subsubsection{Comparison with experimental and numerical results}
The experimental results \cite{legal} obtained for $\eta=0.8$
are in qualitative agreement with a first set of numerical results \cite{rudi}
obtained for $\eta=0.78$. Both studies show that the strato-rotational 
instability occurs for $\mu<\eta$, in agreement with the condition found in 
Eq. (\ref{bsuf}). However, quite recently we have had knowledge of a second set 
of numerical results \cite{rudi2} that do not confirm these previous findings.
The stability line found in \cite{rudi2} is better represented by  $\mu=\mu^*$,
where $\mu^*$ is given explicitly in Eq. (\ref{ci1}). 
We have reported in Table 1 some values of $\mu^*$ corresponding
to values of $\eta$ used in experiments and computations.
In the narrow gap limit, the asymptotic value $\mu^*=1$ is found to agree with the 
stability line for stratified plane Couette flows. 
The computations performed in \cite{rudi2} for three values of the gap size
and different values of the Froude number ($0.5<Fr<2.2$) exhibit 
stability lines $\mu=\mu_s$ with values of $\mu_s$ above the line $\mu=\eta$ 
for $\eta=0.78$ and $\eta=0.5$, in disagreement with numerical and experimental  
results obtained earlier \cite{rudi, legal} for respectively $\eta=0.78$ and 
$\eta=0.8$. For these values of $\eta$,
the stability lines in \cite{rudi2} are in the range $\eta<\mu_s<\mu^*$, in 
reasonable agreement with the results in Eq. (\ref{ci1}).
 
 In the wide gap limit, the necessary instability condition (\ref{ci1}) takes 
the asymptotic form : $\mu<2\eta^2(1+\eta)$,  that fits with
a good accuracy the numerical results found in \cite {rudi2} for $\eta=0.3$. 
The behavior of the numerical neutral curves 
(Reynolds number versus $\mu$) is strongly dependent on the azimuthal
wavenumber \cite{rudi} and on the Froude number \cite{rudi2}. These features 
are not be reproduced by the present analysis based on Eqs. (\ref{Du1c})-(\ref{Dp0c})
which are independent of $l$. Although the Froude number appears in these equations 
through $\widehat m = m Fr$, the instability conditions derived here are independent 
of $\widehat m$.

\begin{table} [phtb] 
\begin{center}
\begin{tabular}{| c |c | c | c|}
\hline 
$\eta$ & $\eta^2$ & $\mu_-$ &$\mu^*$ \\
\hline
0.8  & 0.64   & 0.78  & 0.978\\
0.78 &  0.608 & 0.756 & 0.971 \\
0.5  & 0.25   & 0.4   & 0.625\\
0.3  & 0.09   & 0.165 & 0.231 \\
\hline
\end{tabular} 
\caption{Stability limits for centrifugal instability ($\mu_c=\eta^2$) and
for SRI ($\mu=\mu^*$) as functions of $\eta$. The line $\mu=\mu_-$ 
plays a fundamental role to determine the sign of $t_0$.}
\end{center}
\end {table}

\section{Conclusion}
We performed an inviscid stability analysis of SRI in a Taylor-Couette system 
characterized by a Froude number, $Fr$, measuring the relative importance of  
rotation and stratification. Non axisymmetric disturbances were considered
with azimuthal wavenumber $l$ satisfying $l Fr<<1$ and $l<l_{max}$.

Finite gap effects were taken into account more appropriately than in previous  
inviscid approaches. Although assumptions have been made, they never concerned
the angular velocity profile which is kept equal to $\widehat \Omega=A/r^2+B$. 
Thus, gap size effects manifest themselves through the quantities 
$A$ and $B$ which depend on $\mu$ and $\eta$, the control parameters of 
the system. We derived a necessary instability condition, $\mu<\mu^*$, that fits 
with a good accuracy the recent numerical results of Ref. \cite{rudi2}.
A stronger condition, $\mu<\eta$, found for $\eta>0.38$, better fits with 
earlier experimental and  numerical results \cite{legal, rudi}. For the small 
gap value $\eta=0.3$, the numerical results \cite{rudi, rudi2} exhibit a 
stability line $\mu=\mu_s$ where $\mu_s$ is in good agreement with the asymptotic value
$\mu^* \to 2 \eta^2(1+\eta)$ found for $\eta \to 0$. Unfortunately, in the wide gap limit,
experimental results are still lacking for comparison.
  
The angular velocity profile of circular Couette flow is peculiar since it 
allows an analytical resolution for the pressure perturbations in terms 
of Bessel functions. This is an essential step in the above derivation of the 
stability criterion for SRI in incompressible fluid. A slight change in $\Omega(r)$
can lead to completely different stability results. A flow with constant angular 
momentum ($\Omega\sim r^{-2}$) obtained when $B=0$ in (\ref{CCF}), was considered
in a thin cylindrical shell \cite{papa}. This type of flow is generally assumed 
centrifugally stable, its Rayleigh discriminant being equal to zero.
When $\Phi=0$, Eqs. (\ref{Du1c}) and (\ref{Dp0c}) are decoupled and the above 
analysis for SRI cannot be applied. 
In that case, the existence of nonaxisymmetric unstable modes was proved for 
unstratified flow in a compressible fluid \cite{papa} with an equation of state 
of the type $p \sim \rho^{\gamma}$.
It will be interesting in future work to extend the present analytical
approach to compressible fluids and to other angular velocity profiles.
A first step in this direction was achieved in a recent theoretical approach
\cite{umu2} based on shallow-water approximation for annular 
sections of Keplerian disks. 

{\bf Acknowledgements}\\
The author is grateful to the Referees for their pertinent comments and 
valuable suggestions.
Thanks to P. Le Gal who brought Reference \cite {rudi2} to the author's attention.

\appendix
\renewcommand{\theequation}{\thesection.\arabic{equation}}
\section{Appendix: Limit $l Fr <<1$}
\label{section:A}
\setcounter{equation}{0}
When there is no other restriction on the value of $l$ than $l Fr<<1$, the quantity 
$\bar H=l^2{\bar \sigma}^2-\bar \Phi$ cannot be simplified.
In that case, the governing equations are 
\begin{eqnarray}
\bar \sigma \ D_* \bar u + \frac{\bar Z}{r}  {\bar u} &=& i\ g \ {\bar p} \label{Du}\\
\bar H \bar u - 2 i \frac{\bar \Omega }{r}  \ \bar p &=& -i\ \bar \sigma \ D{\bar p} \label{Dp}
\end{eqnarray}
that depend on the azimuthal wavenumber $l$ through $\bar H$.
Elimination of $\bar u$ gives the equation for $\bar p$
\begin{equation}
\bar \sigma \left[ D^2 \bar p + \left(\frac{1}{r}-\frac{\bar H^{'}}{\bar H}\right) D \bar p\right] 
+ \left[\bar \sigma (\bar m^2 \bar H -\frac{l^2}{r^2})
  + 2 \frac{\bar \Omega}{r} {\partial \over \partial r}
Log\left(\frac{\bar H}{\bar \Omega}\right)\right] \bar p  = 0 \label{D2p0}
\end{equation}
which has a structure analogous to Eq. (\ref{D2p0c}), the main difference is that $\bar H$  
appears instead of $\Phi$. The derivative of the logarythmic term is given by
$$
\frac{\bar H^{'}}{\bar H} - \frac{\bar \Omega^{'}}{\bar \Omega}=-l^2 \bar \sigma \bar \Omega^{'}
{(2\bar \Omega +\bar \sigma) \over \bar H \bar \Omega}
$$
Provided $\bar \sigma \ne 0$, Eq. (\ref{D2p0}) becomes 
\begin{equation}
 D^2 \bar p + \left(\frac{1}{r}-\frac{\bar H^{'}}{\bar H}\right) D \bar p + \left[
 (\bar m^2 \bar H -\frac{l^2}{r^2}) - 2 l^2 \frac{\bar \Omega^{'}}{r\bar H} (2\bar \Omega
  +\bar \sigma)\right]\bar p  = 0 \label{D2p}
\end{equation}
The above equation can be simplified by assuming $\bar m>>l$ or $m>>l Fr^{-1}$. For $l=1$ 
and $Fr=0.5$ this leads to $m>>2$. 
As the critical value of the axial wavenumber is not mentioned in \cite{rudi, rudi2},
it cannot be checked if the assumption is satisfied.
After introducing $\bar p=\bar H^{1/2} \hat p$, one gets the following equation for $\hat p$
\begin{equation}
{\cal L}_0 \hat p+\bar m^2 (l^2{\bar \sigma}^2-\bar \Phi) \hat p=0 \label{pH}
\end{equation}
Eq. (\ref{pH}) is a generalization of Eq. (\ref{L0p}) that takes into account the value of 
the azimuthal wavenumber $l$, its  resolution is left for future work.

{}

\end{document}